\documentclass[aps,prd,twocolumn,preprintnumbers,showpacs,superscriptaddress,nofootinbib,floatfix,10pt]{revtex4-1}
\pdfoutput=1
\usepackage{textcomp}
\usepackage[english]{babel}
\usepackage{amsmath,amssymb,amsbsy,booktabs}
\usepackage{bm}
\usepackage{color}
\usepackage{array}
\usepackage{amstext}
\usepackage{graphicx}
\usepackage{amsfonts}
\usepackage{bm}
\usepackage{dcolumn}
\usepackage{rotating}
\usepackage{epstopdf}
\usepackage{esint}
\usepackage{url}
\usepackage[colorlinks=true,
            linkcolor=blue,
            filecolor=blue,
            anchorcolor=blue,
            urlcolor=blue,
            citecolor=blue
            ]{hyperref}

\usepackage[utf8]{inputenc}

\allowdisplaybreaks[0]

\begin{document}

\title {\bf Muon \boldmath{$g-2$} in a \boldmath{$U(1)$}-symmetric Two-Higgs-Doublet Model}

\author{Shao-Ping Li}
\email{ShowpingLee@mails.ccnu.edu.cn}
\affiliation{Institute of Particle Physics and Key Laboratory of Quark and Lepton Physics~(MOE),\\
Central China Normal University, Wuhan, Hubei 430079, P.~R.~China}

\author{Xin-Qiang Li}
\email{xqli@mail.ccnu.edu.cn}
\affiliation{Institute of Particle Physics and Key Laboratory of Quark and Lepton Physics~(MOE),\\
Central China Normal University, Wuhan, Hubei 430079, P.~R.~China}

\author{Ya-Dong Yang}
\email{yangyd@mail.ccnu.edu.cn}
\affiliation{Institute of Particle Physics and Key Laboratory of Quark and Lepton Physics~(MOE),\\
Central China Normal University, Wuhan, Hubei 430079, P.~R.~China}

\begin{abstract}
\noindent We show in this paper that, in a $U(1)$-symmetric two-Higgs-doublet model (2HDM), the two additional neutral Higgs bosons would become nearly degenerate in the large $\tan\beta$ regime, under the combined constraints from both theoretical arguments and experimental measurements. As a consequence, the excess observed in the anomalous magnetic moment of the muon could not be addressed in the considered framework, following the usual argument where these two neutral scalars are required to manifest a large mass hierarchy. On the other hand, we find that, with an $\mathcal{O}(1)$ top-Yukawa coupling and a relatively light charged Higgs boson, large contributions from the two-loop Barr-Zee type diagrams can account for the muon $g-2$ anomaly at the 1$\sigma$ level, in spite of a large cancellation between the scalar and pseudoscalar contributions. Furthermore, the same scenario can survive the tight constraints from the $B$-physics observables.
\end{abstract}


\maketitle

\section{Introduction}
\label{sec:intro}

There is a long-standing deviation between the standard model (SM) prediction and the experimental measurement for $a_\mu\equiv(g-2)/2$, the anomalous magnetic moment of the muon~\cite{Miller:2007kk,Jegerlehner:2009ry,Miller:2012opa,Lindner:2016bgg,Jegerlehner:2017gek,Aoyama:2020ynm}. The precision measurement of $a_\mu$ has been conducted by the E821 experiment at Brookhaven National Laboratory~\cite{Bennett:2006fi}, the result of which has been recently confirmed by the E989 experiment~\cite{Grange:2015fou} at the Fermi National Laboratory (FNAL) with a smaller uncertainty~\cite{Abi:2021gix}. After combining the two measurements, the current discrepancy between experiment and theory is modified to be~\cite{Abi:2021gix} 
\begin{align}
\Delta a_\mu =a^{\rm exp}_\mu-a^{\rm SM}_\mu=(251\pm59) \times 10^{-11},
\label{delamu}
\end{align}
which increases the discrepancy from $3.7\sigma$ to $4.2\sigma$ and hence strengthens the request for new physics (NP) beyond the SM. Exciting further progress is expected from the FNAL Run-2--4 results and the planned J-PARC $g-2$ experiment~\cite{Abe:2019thb}, as well as from the SM theory~\cite{Aoyama:2020ynm}.

There exist various NP scenarios to explain the muon $g-2$ excess; for recent and thorough reviews, see e.g. Refs.~\cite{Miller:2007kk,Jegerlehner:2009ry,Miller:2012opa,Lindner:2016bgg,Jegerlehner:2017gek,Aoyama:2020ynm}. In this paper, we will consider the two-Higgs-doublet model (2HDM)~\cite{Gunion:1989we,Branco:2011iw}, a simple extension of the SM Higgs sector, as a prospective solution to the muon $g-2$ anomaly. It has been pointed out that the anomaly can be hardly addressed in the type-II 2HDM with a small pseudoscalar Higgs boson mass around $40$ GeV~\cite{Cheung:2003pw,Broggio:2014mna}. This is mainly because the charged Higgs boson mass is now restricted to be larger than $580$ GeV by the branching ratio of the inclusive $\bar{B}\rightarrow X_s\gamma$ decay~\cite{Misiak:2017bgg}, and the resulting mass splitting between the pseudoscalar and the charged Higgs boson would violate the electroweak precision measurements as well as the flavor physics data~\cite{Mahmoudi:2009zx,Deschamps:2009rh,Coleppa:2013dya,Eberhardt:2013uba,Belanger:2013xza,Celis:2013ixa,Chowdhury:2017aav,Haller:2018nnx,Haller:2018nnx}. For the lepton-specific 2HDM, while it can accommodate $\Delta a_\mu$ at the $2\sigma$ level with the parameter space $10\lesssim M_A\lesssim 30$ GeV, $200\lesssim M_{H,H^\pm}\lesssim 350$ GeV and $30 \lesssim \tan\beta\lesssim 50$, the $1\sigma$-level fitted region~\cite{Broggio:2014mna,Wang:2014sda} is already excluded by the lepton universality data in $Z$ and $\tau$ decays~\cite{Abe:2015oca,Chun:2016hzs}. In a general 2HDM with tree-level flavor-changing neutral current (FCNC) at the charged-lepton sector, the muon $g-2$ anomaly can be accounted for with a sizeable $\mu-\tau$ violating coupling~\cite{Omura:2015nja,Omura:2015xcg,Chiang:2016vgf,Iguro:2018qzf}. A reasonable solution to the anomaly can also be provided in the aligned 2HDM~\cite{Han:2015yys,Ilisie:2015tra,Cherchiglia:2016eui,Cherchiglia:2017uwv}. Generically, however, all these scenarios mentioned require a significant mass splitting between the scalar and pseudoscalar Higgs bosons.

In our previous work~\cite{Li:2018rax}, a general 2HDM endowed with electroweak-scale right-handed neutrinos was considered to explain the $B$-physics anomalies $R_{D^{(*)}}$~\cite{Lees:2012xj,Lees:2013uzd,Huschle:2015rga,Aaij:2015yra,Hirose:2016wfn,Sato:2016svk,Hirose:2017dxl,Aaij:2017uff,Aaij:2017deq} and $R_{K^{(*)}}$~\cite{Aaij:2014ora,Aaij:2017vbb}, as well as the neutrino mass problem. We proposed a global $U(1)$ symmetry to induce the sub-eV neutrinos indicated by the neutrino oscillation experiments~\cite{Esteban:2016qun,Capozzi:2017ipn,deSalas:2017kay}. In addition, a large value of $\tan\beta$, the ratio of the two Higgs vacuum expectation values, is favored in light of the $R_{D^{(*)}}$ fits. As will be shown in Sec.~\ref{sec:frame}, in such a large $\tan\beta$ regime, the scalar and pseudoscalar Higgs bosons would become nearly degenerate, as is required by the combined constraints from both theoretical arguments and experimental measurements. As a consequence, an explanation of the muon $g-2$ anomaly in our scenario cannot be realized in the usual way. However, with the particular up-quark FCNC texture specified in Refs.~\cite{Li:2018rax,Crivellin:2015hha}, we will illustrate in this paper that, in spite of the large cancellation between the scalar and pseudoscalar contributions, the $1\sigma$ range of $\Delta a_\mu$ can be accommodated through the two-loop Barr-Zee type diagrams~\cite{Barr:1990vd,Czarnecki:1995wq,Chang:2000ii,Cheung:2001hz,Chen:2001kn,Arhrib:2001xx,Heinemeyer:2003dq,Heinemeyer:2004yq,Abe:2013qla,Ilisie:2015tra,Cherchiglia:2016eui}, in which a sizeable top-quark Yukawa coupling and a relatively light charged Higgs boson are simultaneously presented.

The paper is organized as follows. In Sec.~\ref{sec:frame}, we outline the framework of $U(1)$-symmetric 2HDM and analyze the Higgs mass spectrum. In Sec.~\ref{sec:g-2}, we discuss the dominant contributions to the muon $g-2$ from the two-loop Barr-Zee type diagrams, and then in Sec.~\ref{sec:B-physics} the constraints from the $B$-physics observables are investigated. Our conclusions are finally made in Sec.~\ref{sec:con}. Details of the relevant formulas are relegated to the Appendix.

\section{\texorpdfstring{\boldmath{$U(1)$}}{Lg}-symmetric 2HDM}
\label{sec:frame}

In the 2HDM, the CP-conserving scalar potential with a softly broken $Z_2$ symmetry can be written as~\cite{Branco:2011iw,Gunion:2002zf}
\begin{align}
V&=m_1^2\Phi_1^{\dagger} \Phi_1 +m_2^2 \Phi_2^{\dagger} \Phi_2 -(m_{12}^2\Phi_1^{\dagger} \Phi_2 + \rm{H.c.})
\nonumber \\
&+\dfrac{\lambda_1}{2}(\Phi_1^{\dagger} \Phi_1)^2+ \dfrac{\lambda_2}{2}(\Phi_2^{\dagger} \Phi_2)^2+\lambda_3\Phi_1^{\dagger} \Phi_1\Phi_2^{\dagger} \Phi_2
\nonumber \\
&+\lambda_4\Phi_1^{\dagger} \Phi_2\Phi_2^{\dagger} \Phi_1+\left[ \dfrac{\lambda_5}{2}(\Phi_1^{\dagger} \Phi_2)^2+ {\rm H.c.}  \right],
\label{Higgs potential}
\end{align}
where $\Phi_1$ and $\Phi_2$ are the two Higgs doublets, and $m_{12}^2$ and $\lambda_5$ are real parameters. We will consider the SM-like limit~\cite{Gunion:2002zf,Craig:2013hca,Carena:2013ooa,Dev:2014yca,Bernon:2015qea,Grzadkowski:2018ohf} in which the CP-even neutral scalar $h$ mimics the SM Higgs and its couplings to gauge bosons and fermions attain the corresponding SM values, namely $\beta-\alpha=\pi/2$, where $\alpha$ denotes the mixing angle of the two neutral scalars, and $\tan\beta\equiv v_2/v_1$ with $v_1^2+v_2^2=v^2=(246~{\rm GeV})^2$. In such a limit, the parameters $\lambda_{1-5}$ can be expressed in terms of the physical scalar masses, $\tan\beta$ and $m_{12}^2$ as~\cite{Gunion:2002zf,Kanemura:2004mg}
\begin{align}
\lambda_1v^2&\simeq-m_{12}^2\tan^3\beta+M_H^2\tan^2\beta+M_h^2,
\nonumber \\
\lambda_2v^2&\simeq - \dfrac{m_{12}^2}{\tan\beta}+\dfrac{M_H^2}{\tan^2\beta}+M_h^2,
\nonumber \\
\lambda_3v^2&\simeq -m_{12}^2\tan\beta+2M_{H^{\pm}}^2-M_H^2+M_h^2,
\nonumber \\
\lambda_4v^2&\simeq m_{12}^2\tan\beta+M_A^2-2M_{H^{\pm}}^2,
\nonumber \\
\lambda_5v^2&\simeq m_{12}^2\tan\beta-M_A^2,
\label{lambda form}
\end{align}
which are valid for $\tan\beta\gg 1$. Here, $M_{h,H}$ and $M_A$ are the masses of scalar and pseudoscalar Higgs bosons, respectively, while $M_{H^\pm}$ is the charged Higgs boson mass.

As in our previous work~\cite{Li:2018rax}, we introduce a $U(1)$ symmetry with the charge assignment $L(\Phi_1)=0$ and $L(\Phi_2)=-1$, leading to $\lambda_5=0$, while $m_{12}^2$ is the only soft symmetry breaking source in the Higgs sector. In this case, one can see immediately from Eq.~\eqref{lambda form} that, in the large $\tan\beta\gg 1$ regime, a large mass splitting between $M_{H}$ and $M_A$ would prompt a too large $\lambda_1$, spoiling therefore the validity of perturbativity. Moreover, the perturbativity criterion of $\lambda_{3,4}$ constrains the mass splitting between $M_A$ and $M_{H^\pm}$. On the other hand, the  bounded-from-below conditions (see e.g. Ref.~\cite{Branco:2011iw}) on the scalar potential require $\lambda_{1,2}\geqslant0$, $\lambda_3\geqslant -\sqrt{\lambda_1 \lambda_2}$ and $\lambda_3+\lambda_4-\vert\lambda_5\vert\geqslant-\sqrt{\lambda_1 \lambda_2}$. In addition, the perturbative unitarity of the S-matrix for scalar and longitudinal vector boson scattering puts tight bounds on the Higgs boson masses~\cite{Kanemura:1993hm,Akeroyd:2000wc,Ginzburg:2005dt,Horejsi:2005da,Cacchio:2016qyh}. Finally, the mass spectrum can also have a significant impact on the oblique parameters $S$, $T$ and $U$~\cite{Grimus:2008nb,Haber:2010bw}.

Taking all these constraints into account, we will adopt the perturbativity criteria $\vert \lambda_i \vert\leqslant \pi/2$~\cite{Broggio:2014mna}, the perturbative unitarity bounds $\vert a_{i,\pm}^0\vert\leqslant1/2$, with $a_{i,\pm}^0$ being the eigenvalues of the $s$-wave amplitude matrix for the elastic scattering of two-body bosonic states~\cite{Abe:2015oca}, as well as the $1\sigma$ ranges of $S$, $T$ and $U$ parameters~\cite{Zyla:2020zbs}, with the corresponding formulas collected in the Appendix. In Fig.~\ref{mass bounds}, we show the mass regions allowed in the $(M_A, M_A-M_H)$ plane by these constraints with $\tan\beta=50$, where the blue, red and black boundaries correspond to $M_{H^\pm}=200$, $300$ and $400$ GeV, respectively. One can see clearly from the figure the nearly degenerate pattern in the masses of scalar and pseudoscalar Higgs bosons. Furthermore, the mass of the charged Higgs boson is restricted by the difference $\vert M_{H^\pm}-M_{A,H}\vert\lesssim 100$~GeV. More thorough analyses could also be found e.g. in Refs.~\cite{Bhattacharyya:2013rya,Bhattacharyya:2015nca,Chowdhury:2017aav}.

\begin{figure}[t]
	\centering	
	\includegraphics[width=0.40\textwidth]{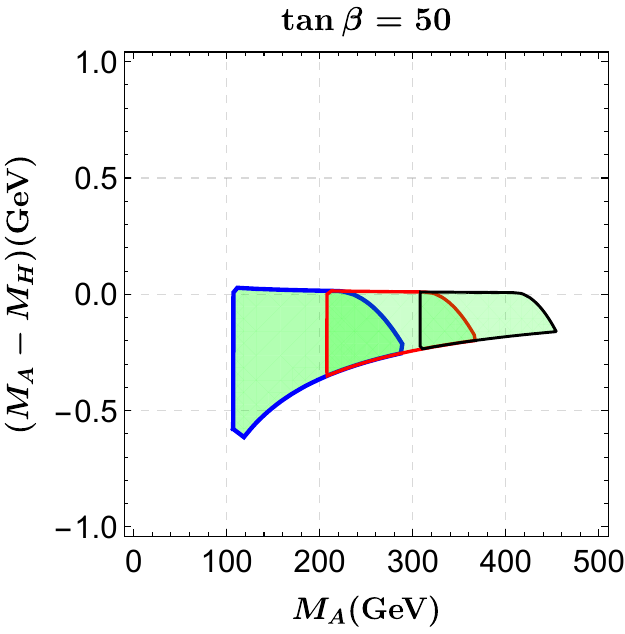}
	\caption{Mass regions allowed in the $(M_A, M_A-M_H)$ plane by the theoretical and experimental constraints (see the text). The blue, red and black boundaries correspond to $M_{H^\pm}=200$, $300$ and $400$ GeV, respectively.}
	\label{mass bounds}
\end{figure}

We should emphasize that, except for the custodial symmetry that protects $M_A\simeq M_{H^\pm}$~\cite{Pomarol:1993mu,Haber:2010bw} or the twisted custodial symmetry that protects $M_H\simeq M_{H^\pm}$~\cite{Gerard:2007kn,Haber:2010bw}, there exists no any symmetry protecting $M_H \simeq M_{A}$. Then, for large tree-level $M_H$ and $M_A$, the radiative self-energy corrections may generate dangerous mass splitting between $M_H$ and $M_A$, and hence violate the constraints under discussion. To this end, we require the (pseudo)scalar masses to be relatively small, e.g., $M_{H,A}\simeq100$~GeV and $M_{H^\pm}\simeq200$~GeV. Such a relatively light charged Higgs boson is also motivated by a feasible explanation for the muon $g-2$ excess under the tight constraints from the $B$-physics observables, as will be discussed in the subsequent sections.

\section{Muon anomalous magnetic moment in the degenerate regime}
\label{sec:g-2}

The $U(1)$ symmetry was also used to constrain the Yukawa coupling textures~\cite{Li:2018rax}. Here we assume that the $U(1)$ symmetry in the quark sector is only an approximate one and allow it to be broken by an additional term in the up-quark sector, treating the coupling $Y_{1}^{u}$ as perturbation~\cite{Crivellin:2015hha} under the following $U(1)$-charge assignment: $L(Q_{Li})=0$, $L(d_{Ri})=1$, $L(u_{Ri})=-1$, $L(\Phi_1)=0$ and $L(\Phi_2)=-1$. Note that we do not include any perturbation in the down-quark sector, as it confronts more severe constraints from flavor physics and collider experiments~\cite{Iguro:2018qzf,Crivellin:2015hha,Crivellin:2013wna,Altunkaynak:2015twa}. As a consequence, the scalar-fermion interaction Lagrangian in our scenario is specified as
\begin{align}
-\mathcal L_{\rm int}&=\bar{Q}_L(Y_1^u \tilde{\Phi}_1+Y_2^u \tilde{\Phi}_2) u_R+ \bar{Q}_L Y^d \Phi_2 d_R
\nonumber \\
&\hspace{-0.15cm} + \bar{E}_LY^\ell \Phi_1 e_R + \bar{E}_L(Y_1^{\nu}\tilde{\Phi}_1+Y_2^{\nu} \tilde{\Phi}_2)N_R+ {\rm H.c.},
\label{lag}
\end{align}
where $\tilde{\Phi}_i=i\tau_2 \Phi_i^\ast$ with $\tau_2$ being the Pauli matrix; $Q_L$, $E_L$, $u_R$, $d_R$ and $e_R$ denote the left-handed quark and lepton doublets, the right-handed up-quark, down-quark, and charged-lepton singlets, respectively. The right-handed neutrino singlet $N_R$ was introduced to account for the small neutrino mass and the $R_{K^{(*)}}$ anomaly~\cite{Li:2018rax}. Note that similar frameworks relevant to the muon $g-2$ are discussed in Refs.~\cite{Chiang:2018bnu,Chiang:2017vcl,Campos:2017dgc}.

In this paper, we generalize the previously studied texture of the up-quark Yukawa coupling to\footnote{In this case, for $\tan\beta=\mathcal{O}(50)$ and $\vert \epsilon_{ct,tc,tt}\vert \leqslant \mathcal{O}(1)$, the quark masses and mixings can be reproduced without a significant degree of fine-tuning~\cite{Iguro:2018qzf,Crivellin:2015hha,Crivellin:2013wna}. At the same time, the particular texture of Eq.~\eqref{FCNC repre}, together with the favored parameter regions of the corresponding entries, provides a phenomenologically viable scenario for our purpose.}
\begin{align}
X^u_1\equiv \dfrac{1}{\sqrt{2}}V_L^u Y_1^u V_R^{u \dagger}=\left(\begin{array}{ccc}
  0 & 0 &0 \\
  0 & 0 &\epsilon_{ct} \\
  0 & \epsilon_{tc} &\epsilon_{tt}
\end{array}\right),
\label{FCNC repre}
\end{align}
where $V_{L,R}^u$ denote the basis transformation matrices in the up-quark sector. The Yukawa texture in Eq.~\eqref{FCNC repre} was studied in the general 2HDM, namely the type-III 2HDM, with or without the Cheng-Sher ansatz~\cite{Iguro:2018qzf,Li:2018rax,Crivellin:2015hha,Kim:2015zla,Wang:2016ggf,Iguro:2017ysu,Altunkaynak:2015twa,Crivellin:2013wna,Chen:2018hqy,Cline:2015lqp}. It has been pointed out that a nonzero $\epsilon_{tc}$ can improve the discrepancy observed in $R_{D^{(*)}}$~\cite{Crivellin:2015hha,Iguro:2017ysu,Li:2018rax}, while $\mathcal{O}(1)$ $\epsilon_{tt}$ is required to explain the $R_{K^{(*)}}$ anomaly~\cite{Li:2018rax}. It will be shown in this section that $\mathcal{O}(1)$ $\epsilon_{tt}$ is also responsible for resolving the muon $g-2$ anomaly. On the other hand, the large contributions from $\epsilon_{tt}$ to $\bar{B}\rightarrow X_s\gamma$ and $B_s-\bar{B}_s$ mixing can be cancelled to a large extent, if a nonzero $\epsilon_{ct}$ is presented at the same time~\cite{Crivellin:2013wna,Altunkaynak:2015twa,Iguro:2018qzf,Chen:2018hqy}, as will be discussed in Sec.~\ref{sec:B-physics}.

\begin{figure}[t]
	\centering	
	\includegraphics[width=0.40\textwidth]{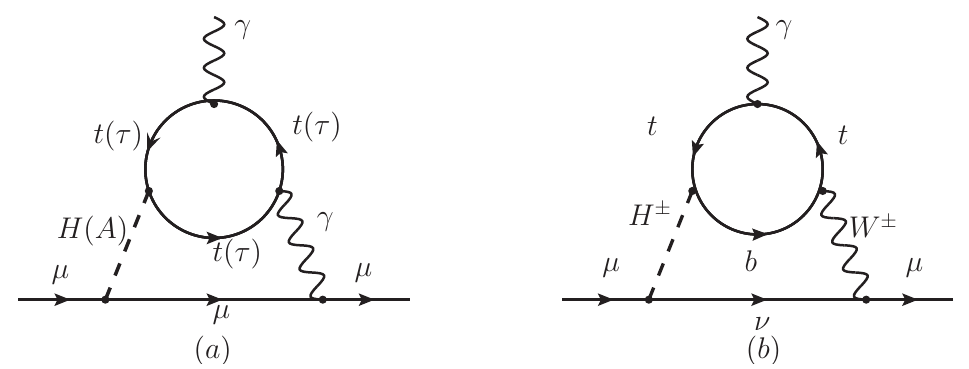}
	\caption{$(a)$: typical two-loop Barr-Zee type diagrams with the top ($\tau$) loop. $(b)$: new sets of two-loop Barr-Zee type diagrams with the top-bottom (bottom-top) loop.}
	\label{BZs}
\end{figure}

In our scenario, the two crucial elements for addressing the muon $g-2$ excess are: (i) a sizeable $\epsilon_{tt}$ and (ii) a relatively light charged Higgs boson. They provide the dominant contributions to the muon $g-2$ via the typical two-loop Barr-Zee type diagrams shown in Fig.~\ref{BZs}$(a)$~\cite{Barr:1990vd,Czarnecki:1995wq,Chang:2000ii,Cheung:2001hz,Chen:2001kn,Arhrib:2001xx,Heinemeyer:2003dq,Heinemeyer:2004yq,Abe:2013qla,Ilisie:2015tra,Cherchiglia:2016eui} as well as the new sets of two-loop Barr-Zee type diagrams shown in Fig.~\ref{BZs}$(b)$~\cite{Ilisie:2015tra}. For Fig.~\ref{BZs}$(a)$, the fermion loops come from the top quark and the $\tau$ lepton, while the bottom-quark loop has a suppression factor $1/\tan\beta$ in the large $\tan\beta$ regime  and there is no neutrino loop. Note that the amplitude  with the photon propagator replaced by that of the $Z$ boson is suppressed by a factor $-1/4+\sin^2\theta_W\approx -0.02$~\cite{Chang:2000ii}.

As for the new sets of two-loop Barr-Zee type diagrams discussed in Ref.~\cite{Ilisie:2015tra}, we would like to make the following remarks. Due to the assumption of CP conservation in the scalar potential, there are neither $A H^\pm H^\mp$ nor $A W^\pm W^\mp$ couplings. Then, contributions from all the diagrams involving these vertices shown in Fig.~\ref{BZs} would vanish. Meanwhile, the contribution involving the $H H^\pm H^\mp$ vertex is suppressed in the large $\tan\beta$ regime, because the $H H^\pm H^\mp$ coupling is proportional to $1/\tan\beta$ (we have confirmed this point with the code \texttt{SARAH}~\cite{Staub:2013tta}). Finally, due to the absence of the $H W^\pm W^\mp$ coupling in the SM-like limit, there exist no contributions from the diagrams involving this vertex either. Therefore, in our scenario, the diagrams shown in Fig.~\ref{BZs}$(b)$ with the top-bottom (bottom-top) loop should give the dominant contributions among all these two-loop Barr-Zee type diagrams. We should also mention that, although the heavy neutrino with mass around the electroweak scale is introduced in our framework, we need not consider its contribution coming from Fig.~\ref{BZs}$(b)$ with the light neutrino propagator replaced by that of the heavy neutrino $N$, because the $\mathcal{O}(1)$ $H^\pm N \mu^\mp$ coupling considered in Ref.~\cite{Li:2018rax} is compensated by the small $W^\pm N \mu^\mp$ one~\cite{Antusch:2006vwa,Akhmedov:2013hec,Fernandez-Martinez:2015hxa,Fernandez-Martinez:2016lgt}.

\begin{figure}[t]
	\centering
	\includegraphics[width=0.40\textwidth]{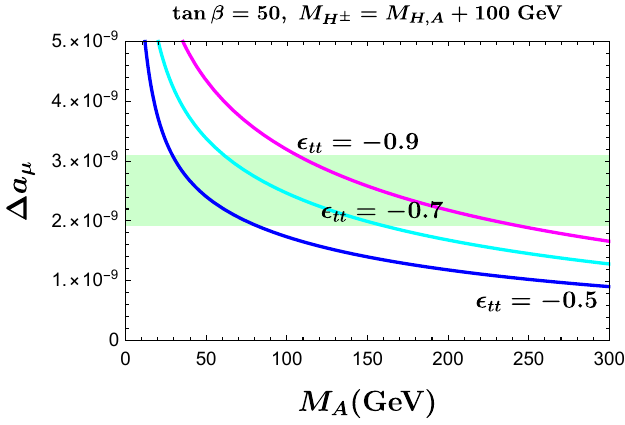}
	\caption{Dependence of $\Delta a_\mu$ on the pseudoscalar mass $M_A$ with different values of the top-Yukawa coupling $\epsilon_{tt}$. The green band corresponds to the $1\sigma$ range of $\Delta a_\mu$ given by Eq.~\eqref{delamu}.}
	\label{amu}
\end{figure}

Based on these observations, we will only consider the two-loop Barr-Zee type diagrams shown in Fig.~\ref{BZs}. The corresponding amplitudes as well as the one-loop contributions generated by the $H$, $A$ and $H^\pm$ propagators are collected in the Appendix. In the numerical analysis, we fix $\tan\beta=50$ and $M_{H^\pm}= M_{H,A}+100~{\rm GeV}$ which is allowed by the mass spectrum shown in Fig.~\ref{mass bounds}. In addition, the heavy neutrino mass is fixed at $200$ GeV and an $\mathcal{O}(1)$ neutrino Yukawa coupling is adopted~\cite{Li:2018rax} (see the Appendix). The total NP contributions to the muon $g-2$ are shown in Fig.~\ref{amu}, where the dependence of $\Delta a_\mu$ on the pseudoscalar mass $M_A$ with different values of the top-Yukawa coupling $\epsilon_{tt}$ is displayed. It can be seen that the $1\sigma$ range of $\Delta a_\mu$ given by Eq.~\eqref{delamu} can be accommodated depending on the values of the top-Yukawa coupling $\epsilon_{tt}$ and the (pseudo)scalar mass $M_{H(A)}$. For $\epsilon_{tt}=-0.5$, we obtain $30 \lesssim M_{H,A}\lesssim 80$ GeV, while for $\epsilon_{tt}=-0.7$, the allowed mass region increases to $60\lesssim M_{H,A}\lesssim 160$ GeV; however, with $\epsilon_{tt}=-0.9$ and $M_{H,A}\lesssim 110$ GeV, the resulting $\Delta a_\mu$ would exceed the $1\sigma$ range.

It should also be mentioned that the two-loop Barr-Zee type diagrams shown in Fig.~\ref{BZs}$(b)$ can also give a contribution to the radiative decay $\mu \rightarrow e\gamma$ if the final lepton is replaced with an electron. However, the amplitude would carry an additional factor $U_{\mu j}^\nu U_{ej}^{\nu *}$, where $U^\nu$ represents the full $6\times 6$ neutrino mixing matrix, in the convention specified in Ref.~\cite{Li:2018rax}. Therefore, the amplitude would be proportional to $\sum_{j=1}^3 U_{\mu j+1}^\nu U_{ej+1}^{\nu *}\equiv 2\eta_{\mu e}$, where $\eta$ represents the non-unitary effect of the neutrino mixing matrix~\cite{Fernandez-Martinez:2016lgt}. Following the effective Lagrangian method~\cite{Crivellin:2015hha} and taking the  $2\sigma$ upper bound, $\vert 2\eta_{\mu e}\vert \lesssim \mathcal{O}(10^{-5})$~\cite{Fernandez-Martinez:2016lgt}, we would obtain $\mathcal{B}(\mu \rightarrow e \gamma)\lesssim \mathcal{O}(10^{-15})$,  which is two orders of magnitude smaller than the current bound, $4.2\times 10^{-13}$~\cite{TheMEG:2016wtm}. We can, therefore, conclude that our explanation for the muon $g-2$ discrepancy does not conflict with the stringent constraint from $\mu \rightarrow e \gamma$.

\section{Constraints from \texorpdfstring{\boldmath{$B$}}{Lg}-physics observables}
\label{sec:B-physics}

In this section, we analyze the tight constraints from the $B$-physics observables, concentrating on the branching ratio of the inclusive $\bar{B}\rightarrow X_s\gamma$ decay and the mass difference in the $B_s-\bar{B}_s$ mixing, as the sizeable top-Yukawa coupling $\epsilon_{tt}$ and the relatively small charged Higgs boson mass give large contributions to these observables~\cite{Crivellin:2013wna,Altunkaynak:2015twa,Iguro:2018qzf,Chen:2018hqy}.

\subsection{\texorpdfstring{\boldmath{$\bar{B}\rightarrow X_s\gamma$}}{Lg}}

The low-energy effective Hamiltonian for the inclusive $\bar{B}\rightarrow X_s\gamma$ decay is given by
\begin{align}
\mathcal{H}_{\rm eff}^{b\rightarrow s\gamma}=-\dfrac{4G_F}{\sqrt{2}}V_{ts}^* V_{tb} \sum_{i=1}^8 C_i(\mu)\mathcal{O}_i(\mu),
\label{bsgaeffH}
\end{align}
where the current-current operators $\mathcal{O}_{1,2}$ and the QCD-penguin operators $\mathcal{O}_{3-6}$ could be found e.g. in Ref.~\cite{Borzumati:1998tg}, while the dipole operators $\mathcal{O}_{7\gamma}$ and $\mathcal{O}_{8g}$ are defined, respectively, as
\begin{align}
\mathcal{O}_{7\gamma}&=\dfrac{e}{16\pi^2}m_b\bar{s}\sigma^{\mu \nu} P_{R} b F_{\mu\nu},
\nonumber \\[0.2cm]
\mathcal{O}_{8g}&=\dfrac{g_s}{16\pi^2}m_b\bar{s}\sigma^{\mu \nu} P_{R} T^a  b G^a_{\mu\nu}.
\label{eq:dipole}
\end{align}
Note that the primed operators $\mathcal{O}^\prime_{7\gamma,8g}$ that are obtained from Eq.~\eqref{eq:dipole} by
replacing $P_R$ with $P_L$ need not be included in Eq.~\eqref{bsgaeffH}, because the Wilson coefficients of these operators are suppressed by $m_s/m_b$ relative to those coming from $\mathcal{O}_{7\gamma,8g}$ in the SM, and are zero in our scenario due to the absence of FCNCs in the down-quark sector.

Following Ref.~\cite{Buras:2011zb}, the branching ratio of the inclusive $\bar{B}\rightarrow X_s\gamma$ decay can be expressed as
\begin{align}
\mathcal{B}(\bar{B}\rightarrow X_s \gamma)=R\left[\vert C_{7\gamma}(\mu_b)\vert^2+N(E_\gamma) \right],
\end{align}
where $R$ is an overall factor and determined to be $R=2.47\times10^{-3}$~\cite{Gambino:2001ew,Misiak:2006ab}, while $N(E_\gamma)$ denotes the non-perturbative correction, with $N(E_\gamma)=(3.6\pm0.6)\times10^{-3}$~\cite{Gambino:2001ew} for a photon-energy cut off $E_\gamma>1.6$ GeV in the $\bar{B}$-meson rest frame. The Wilson coefficient $C_{7\gamma}(\mu_b)$ can be decomposed into the sum of the SM and the NP contributions:
\begin{equation}
C_{7\gamma}(\mu_b)=C^{\rm SM}_{7\gamma}(\mu_b)+C^{\rm NP}_{7\gamma}(\mu_b).
\end{equation}
For the SM part, we adopt the result at the next-to-next-to leading order in QCD which gives $\mathcal{B}_{s\gamma}^{\rm SM}=(3.36\pm0.23)\times 10^{-4}$  for $E_\gamma> 1.6$ GeV~\cite{Czakon:2015exa,Misiak:2015xwa}, while for the NP contribution, we use the leading order result at the scale $\mu_b=\mathcal{O} (m_b)$:
\begin{equation}
C^{\rm NP}_{7\gamma}(\mu_b)= \kappa_7 C^{\rm NP}_{7\gamma}(\mu_H)+\kappa_8 C^{\rm NP}_{8g}(\mu_H),
\end{equation}
where the Wilson coefficients $C^{\rm NP}_{7\gamma}(\mu_H)$ and $C^{\rm NP}_{8g}(\mu_H)$ at the initial scale $\mu_H= \mathcal{O} (M_{H^{\pm}})$ are collected in the Appendix. For the numerical study, we take the magic numbers $\kappa_7=0.524$ and $\kappa_8=0.118$ evaluated at $\mu_H=200$ GeV and $\mu_b=2.5$ GeV~\cite{Buras:2011zb}.

\subsection{\texorpdfstring{\boldmath{$B_s-\bar{B}_s$}}{Lg} mixing}

Adopting the overall normalization of the SM contribution, the low-energy effective Hamiltonian for $B_s-\bar{B}_s$ mixing can be written as
\begin{align}
\mathcal{H}_{\rm eff}^{\Delta B=2}= \dfrac{G_F^2 (V_{tb}^* V_{ts})^2}{16\pi^2}M_W^2\left(\sum_{i=1}^5 C_i \mathcal{O}_i+\sum_{i=1}^3 \tilde{C}_i \tilde{\mathcal{O}}_i\right),
\end{align}
where the four-quark operators could be found e.g. in Refs.~\cite{Buras:2001ra,Becirevic:2001jj}. In our scenario, only the Wilson coefficient $C_1$ of the operator
\begin{align}
\mathcal{O}_1=\left(\bar{b}^\alpha \gamma_\mu P_L s^\alpha \right) \left(\bar{b}^\beta \gamma^\mu P_L s^\beta \right)
\end{align}
receives a significant NP contribution, while the NP contributions to the other Wilson coefficients are either absent as no FCNCs occur in the down-quark sector or suppressed by $1/\tan\beta$ in the large $\tan\beta$ regime. This can also be understood e.g. from Refs.~\cite{Crivellin:2013wna,Chen:2018hqy}.

A prominent constraint from the $B_s-\bar{B}_s$ mixing is due to the mass difference between the two mass eigenstates of the
neutral $B_s$ mesons which is given as
$\Delta M_s=2\vert M^s_{12}\vert$, where $M_{12}^s$ is the off-diagonal element in the $B_s$-meson mass matrix, with the known SM result given by~\cite{Ball:2006xx,Artuso:2015swg,DiLuzio:2017fdq}
\begin{align}
M^{s,\rm SM}_{12}=\dfrac{G_F^2}{12\pi^2}(V_{tb}^*V_{ts})^2M_W^2M_{B_s} f_{B_s}^2 \hat{\eta}_B B S_0(x_t),
\end{align}
where $M_{B_s}$ and $f_{B_s}$ are the $B_s$-meson mass and decay constant, respectively. The factor $\hat{\eta}_B$ encodes the short-distance QCD corrections~\cite{Buras:1990fn}, while the bag parameter $B$~\cite{Bazavov:2016nty}, together with the decay constant $f_{B_s}$, parameterizes all the long-distance QCD effect contained in the hadronic matrix element $\langle \bar{B}_s|\mathcal{O}_1|B_s\rangle$. Note that a different convention for the QCD correction and the bag parameter is also used in the literature through the relation $\hat{\eta}_B B\equiv \eta_B \hat{B}$ (see e.g. Ref.~\cite{Aoki:2016frl} and the relevant discussions in Refs.~\cite{Lenz:2010gu,DiLuzio:2017fdq}). The Inami-Lim function $S_0(x_t\equiv\left(\overline{m}_t(\overline{m}_t)\right)^2/M_W^2)\approx2.39$~\cite{Inami:1980fz} is calculated from the $W$-boson box diagrams with two internal top-quark exchanges, the value of which is obtained with the central value of the top-quark $\overline{\rm MS}$ mass $\overline{m}_t(\overline{m}_t)$~\cite{TevatronElectroweakWorkingGroup:2016lid}.

Normalizing to the SM contribution, we can express $\Delta M_s$ in the presence of NP contributions as
\begin{equation}
\Delta M_s=\Delta M_s^{\rm SM} \vert 1+\Delta_s^{\rm NP}\vert,
\label{deltaMsp}
\end{equation}
where $\Delta M_s^{\rm SM}=(18.4^{+0.7}_{-1.2})~{\rm ps}^{-1}$ is taken from the 2019 updated result presented in Refs.~\cite{DiLuzio:2019jyq,Lenz:2019lvd} and $\Delta_s^{\rm NP}=C_1^{\rm NP}(\mu_H)/(4S_0(x_t))$. The Wilson coefficient $C_1^{\rm NP}(\mu_H)$ evaluated at $\mu_H=\mathcal{O}(M_{H^\pm})$ is given in the Appendix. It should be mentioned that, in obtaining Eq.~\eqref{deltaMsp}, we have assumed that the renormalization group running effect on the NP contributions~\cite{Buras:2001ra,Becirevic:2001jj} from the initial scale $\mu_H$ down to the low scale $\mu_b$ is identical to that in the SM part (which is encoded in the factor $\hat{\eta}_B$). This approximation is quite reasonable, because the charged Higgs boson mass is considered to be around the electroweak scale, being not far from the top-quark mass.

\subsection{Combined constraints from \texorpdfstring{\boldmath{$\mathcal{B}(\bar{B}\rightarrow X_s \gamma)$}}{Lg} and \texorpdfstring{\boldmath{$\Delta M_s$}}{Lg}}

\begin{figure}[t]
	\centering
	\includegraphics[width=0.40\textwidth]{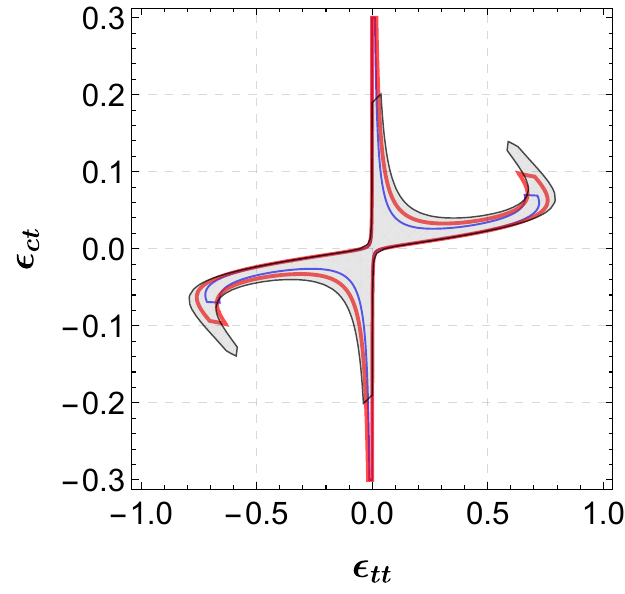}
	\caption{Allowed parameter space in the $(\epsilon_{tt},\epsilon_{ct})$ plane under the combined constraints from $\bar{B}\rightarrow X_s \gamma$ decay and $B_s-\bar{B}_s$ mixing. The regions with blue, red and black boundaries correspond to $M_{H^\pm}=200$, $300$ and $400$ GeV, respectively.}
	\label{B-fits}
\end{figure}

We now proceed to analyze the combined constraints from the branching ratio $\mathcal{B}(\bar{B}\rightarrow X_s \gamma)$ and the mass difference $\Delta M_s$. To this end, we confront our theoretical predictions with the $1\sigma$ ranges of the world-averaged results compiled by the Heavy Flavor Averaging Group for these two observables~\cite{Amhis:2016xyh}: $\mathcal{B}_{s\gamma}^{\rm exp}=(3.32\pm0.15)\times 10^{-4}$ for $E_\gamma
>1.6$ GeV and $\Delta M_s^{\rm exp}=(17.757\pm0.021)~ \rm ps^{-1}$. In Fig.~\ref{B-fits}, by fixing $\epsilon_{tc}=0.1$ in light of the combined fits for $R_{D^{(*)}}$ and $B_c^-\rightarrow \tau^- \bar{\nu}$~\cite{Iguro:2017ysu,Li:2018rax}, we show the allowed parameter space in the $(\epsilon_{tt},\epsilon_{ct})$ plane, with $M_{H^\pm}=200$, $300$ and $400$ GeV, corresponding to the blue, red and black boundaries, respectively. We can see clearly from this figure that it is possible to allow $\epsilon_{tt}=-0.7$ and $M_{H^\pm}=200$ GeV even with a tightly constrained $\epsilon_{ct}$: $-0.04\lesssim \epsilon_{ct}\lesssim -0.07$. It is therefore concluded that the parameter region with large $\epsilon_{tt}$ and small $M_{H^\pm}$ required by the muon $g-2$ excess can still be reached under the tight constraints from these two observables.

Finally, let us discuss briefly the compatibility of our scenario with the direct collider constraints. In a framework similar to what is considered here, it has been found that the neutral and charged Higgs bosons with masses being as light as 200 GeV are still compatible with the current LHC constraints~\cite{Gori:2017tvg,Hou:2018zmg}. For a pseudoscalar boson being lighter than the top quark, it is also found that the constraints from the top-quark decay and the same-sign top pair production at the LHC still allow $M_A\simeq150$ GeV, while the 13 TeV LHC direct constraints on the charged Higgs boson mass would be weaker than those from the indirect $B$-meson physics~\cite{Wang:2016ggf}. More detailed analysis with scalar masses being around 200 GeV can also be found e.g. in Ref.~\cite{Iguro:2017ysu}. Regarding the LEP constraints, the current bound on the charged Higgs boson mass is $M_{H^{\pm}}>80$ GeV~\cite{Abbiendi:2013hk}, while the $95\%$ C.L. lower bound of the neutral scalars is 95 GeV in the limit where $M_H\simeq M_A$~\cite{Schael:2006cr,Cline:2015lqp}. Based on the observations, we can, therefore, conclude that the preferred region found in this paper is compatible with both the LHC and LEP constraints.

\section{Conclusions}
\label{sec:con}

In this paper, we have demonstrated that in the $U(1)$-symmetric 2HDM, where the two Higgs doublets carry different global $U(1)$ charges, the two additional neutral Higgs bosons would become nearly degenerate in the large $\tan\beta$ regime, due to the constraints from both theoretical arguments and experimental measurements. As a result, it is impossible to address the muon $g-2$ anomaly in this framework, because a significant mass splitting between the scalar and pseudoscalar Higgs bosons, as is generally required in the usual 2HDMs, cannot be realized. However, with an $\mathcal{O}(1)$ top-Yukawa coupling and a relatively light charged Higgs boson mass ($\simeq200$ GeV), we found that there exist large contributions to the muon $g-2$ arising from the two-loop Barr-Zee type diagrams, in spite of a large cancellation between the scalar and pseudoscalar sectors, providing therefore an explanation for the muon $g-2$ excess at the $1\sigma$ level. At the same time, the required top-Yukawa coupling and charged Higgs boson mass can also survive the tight constraints from the branching ratio of the inclusive $\bar{B}\rightarrow X_s \gamma$ decay and the mass difference in the $B_s-\bar{B}_s$ mixing.

\section*{Acknowledgements}

This work is supported by the National Natural Science Foundation of China (Grant Nos.~11675061, 11775092, 11521064 and 11435003). X.~Li is also supported in part by the self-determined research funds of CCNU from the colleges' basic research and operation of MOE~(CCNU18TS029).

\section*{Appendix}

In this Appendix, we collect the relevant expressions for the oblique parameters $S$, $T$ and $U$, the well-known one-loop and the two-loop Barr-Zee type contributions to the muon $g-2$, as well as the relevant Wilson coefficients in the inclusive $\bar{B}\rightarrow X_s\gamma$ decay and the $B_s-\bar{B}_s$ mixing, in the framework of $U(1)$-symmetric 2HDM.

\subsection{Oblique parameters}

Following Refs.~\cite{Grimus:2008nb,Haber:2010bw}, the oblique parameters $S$, $T$ and $U$ in the SM-like limit are given, respectively, by
\begin{align}
&S=\frac{\left[\mathcal{B}_{22}(M_Z^2,M_H^2,M_A^2)-\mathcal{B}_{22}(M_Z^2,M_{H^{\pm}}^2,M_{H^{\pm}}^2)\right]}{\pi M_Z^2},\\
&T=\dfrac{1}{16\pi \sin^2\theta_W M_W^2} \left[ F\left(M_{H^\pm}^2,M_H^2\right)+ F\left(M_{H^\pm}^2,M_A^2\right)\right.
\nonumber \\
&\hspace{3.0cm}-\left. F\left(M_H^2,M_A^2\right)\right],
 \\
&S+U=\dfrac{1}{\pi M_W^2}\left[\mathcal{B}_{22}(M_W^2,M_{H^{\pm}}^2,M_A^2)
\right.
\nonumber \\
&+\left. \mathcal{B}_{22}(M_W^2,M_{H^{\pm}}^2,M_H^2)-2\mathcal{B}_{22}(M_W^2,M_{H^{\pm}}^2,M_{H^{\pm}}^2)\right],
\label{STU}
\end{align}
with
\begin{align}
\mathcal{B}_{22}(q^2,m_1^2,m_2^2)&=B_{22}(q^2,m_1^2,m_2^2)-B_{22}(0,m_1^2,m_2^2),
\\
B_{22}(q^2,m_1^2,m_2^2)&=\frac{1}{4}(\Delta+1)\left(m_1^2+m_2^2-\frac{1}{3}q^2\right)
\nonumber \\
&-\frac{1}{2}\int^1_0 dx X \ln(X-i\epsilon),
\\[0.2cm]
F(x,y)&=\dfrac{x+y}{2}-\dfrac{x y}{x-y}\ln\left(\dfrac{x}{y}\right),
\end{align}
where $X\equiv m_1^2 x+m_2^2(1-x)-q^2 x (1-x)$, and $\Delta\equiv\frac{2}{4-d}+\ln(4\pi)-\gamma$ in $d$-dimensional space-time.

\subsection{Formulas for muon \texorpdfstring{\boldmath{$g-2$}}{Lg}}

The amplitude for the typical two-loop Barr-Zee type diagrams shown in Fig.~\ref{BZs}$(a)$ is given by~\cite{Barr:1990vd,Czarnecki:1995wq,Chang:2000ii,Cheung:2001hz,Chen:2001kn,Arhrib:2001xx,Heinemeyer:2003dq,Heinemeyer:2004yq,Abe:2013qla,Ilisie:2015tra,Cherchiglia:2016eui}
\begin{align}
\Delta a_\mu^{(a)}&=\dfrac{G_F \,m_\mu^2}{4\sqrt{2}\pi^2} \dfrac{\alpha}{\pi}\sum_{i,f}N_c^f Q_f^2\, r_f^i\,y_\mu^i\, y_f^i\, g_i(r_f^i),
\end{align}
with
\begin{align}
g_i(r_f^i)=\int_0^1 dx \dfrac{\mathcal{N}_i(x)}{x(1-x)-r_f^i} \ln\dfrac{x(1-x)}{r_f^i},
\end{align}
where $G_F$ is the Fermi constant and $\alpha$ the fine-structure constant; $r_f^i=m_f^2/M_i^2$ with $i=H,\,A$, and $m_f$, $Q_f$ and $N_c^f$ are the mass, the electric charge (in unit of the elementary charge), and the number of colors for fermion $f=t,\,\tau$, respectively; $\mathcal{N}_H(x)=2x(1-x)-1$ and $\mathcal{N}_A(x)=1$. In our scenario, the fermion couplings of neutral Higgs bosons are given by $y_{\mu,\tau}^H=-y_{\mu,\tau}^A=\tan\beta$ and $y_t^H=y_t^A=\dfrac{v}{m_t}\epsilon_{tt}$.

The amplitude for the new sets of two-loop Barr-Zee type diagrams shown in Fig.~\ref{BZs}$(b)$ is given by~\cite{Ilisie:2015tra,Crivellin:2015hha}
\begin{align}
\Delta a_\mu^{(b)}&=\dfrac{\alpha \,m_\mu^2\, N_c \,\vert V_{tb}\vert^2}{32\pi^3 \sin^2\theta_W (M_{H^\pm}^2-M_W^2)}\tan\beta \dfrac{m_t}{v}\epsilon_{tt}
\nonumber \\
&\times \int_0^1 \left[ Q_t x+Q_b(1-x)\right] x (1+x)
\nonumber \\
&\times \left[\mathcal{G}\left(\dfrac{m_t^2}{M_{H^\pm}^2},\dfrac{m_b^2}{M_{H^\pm}^2}\right)-\mathcal{G}\left(\dfrac{m_t^2}{M_{W}^2}, \dfrac{m_b^2}{M_{W}^2}\right)\right],
\end{align}
where $V_{tb}$ is the CKM matrix element, and
\begin{align}
\mathcal{G}\left(r_a, r_b\right)=\dfrac{\ln\left(\dfrac{r_a x+r_b (1-x)}{x(1-x)}\right)}{x(1-x)-r_a x-r_b(1-x)}.
\end{align}

The well-known one-loop amplitudes generated by the neutral and charged Higgs boson propagators are
given by~\cite{Barr:1990vd,Czarnecki:1995wq,Chang:2000ii,Cheung:2001hz,Chen:2001kn,Arhrib:2001xx,Heinemeyer:2003dq,Heinemeyer:2004yq,Abe:2013qla,Ilisie:2015tra,Cherchiglia:2016eui}
\begin{align}
\Delta a_\mu^{H,A}&=\dfrac{G_F\, m_\mu^2}{4\sqrt{2}\pi^2}\tan^2\beta \left[\dfrac{m_\mu^2}{M_H^2}\left(-\ln\dfrac{m_\mu^2}{M_H^2}-\dfrac{7}{6}\right)\right.
\nonumber \\
&+\left.\dfrac{m_\mu^2}{M_A^2}\left(\ln\dfrac{m_\mu^2}{M_A^2}+\dfrac{11}{6}\right)\right],
\\[0.2cm]
\Delta a_\mu^{H^\pm}&=\dfrac{G_F \,m_\mu^2}{4\sqrt{2}\pi^2}\tan^2\beta \dfrac{m_\mu^2}{M_{H^\pm}^2} \left(-\dfrac{1}{6}\right)
\nonumber \\
&+\dfrac{\vert x_2\vert^2}{16\pi^2}\dfrac{m_\mu^2}{M_{H^\pm}^2} \left(-\dfrac{1}{6}\right)\mathcal{J}\left( \dfrac{M^2}{M_{H^\pm}^2}\right),
\end{align}
with
\begin{align}
\mathcal{J}(x)&=\dfrac{2x^3+3x^2-6x^2\ln x-6x+1}{(x-1)^4}.
\end{align}
Note that in obtaining $\Delta a_\mu^{H,A}$, we have neglected the $\mathcal{O}(m_\mu^4/M_{H,A}^4)$ terms. The first term in $\Delta a_\mu^{H^\pm}$ stems from the diagram with light neutrino propagator, while the second term from the heavy neutrino propagator~\cite{Ma:2001mr}. In line with our previous work~\cite{Li:2018rax}, we take the muon-philic coupling $x_2=1$ and the heavy neutrino mass $M=200$ GeV during the numerical analysis.

\subsection{Wilson coefficients in \texorpdfstring{\boldmath{$\bar{B}\rightarrow X_s\gamma$}}{Lg} decay and \texorpdfstring{\boldmath{$B_s-\bar{B}_s$}}{Lg} mixing}

For the inclusive  $\bar{B}\rightarrow X_s\gamma$ decay, in our scenario, only $C_{7\gamma}^{\rm NP}(\mu_H)$ and $C_{8g}^{\rm NP}(\mu_H)$ get a significant NP contribution from the one-loop penguin diagram with charged Higgs boson exchange. They are given as~\cite{Crivellin:2013wna,Altunkaynak:2015twa,Iguro:2018qzf,Chen:2018hqy}
\begin{align}
C_{7\gamma,8g}^{\rm NP}(\mu_H)&=\dfrac{v^2}{m_t^2} \left(\dfrac{V_{cs}^*}{V_{ts}^*}\epsilon_{ct}\epsilon_{tt}+\vert\epsilon_{tt}\vert^2\right)E_{7,8}(y_t)
\nonumber \\
&+\dfrac{v^2}{m_c^2}\vert\epsilon_{tc}\vert^2 E_{7,8}(y_c),
\end{align}
where $y_{t,c}\equiv m_{t,c}^2/M^2_{H^\pm}$, and the scalar functions $E_{7,8}$ are defined by
\begin{align}
E_7(y)&=\frac{y\left[\left(18 y^2-12 y\right)\ln y - 8y^3+3 y^2+12 y-7\right]}{72 (y-1)^4},
\nonumber \\[0.15cm]
E_{8}(y)&=\frac{y \left(-6 y \ln y -y^3+6 y^2-3 y-2\right)}{24 (y-1)^4},
\end{align}
which can be found e.g. in Refs.~\cite{Borzumati:1998tg,Crivellin:2013wna,Altunkaynak:2015twa,Iguro:2018qzf,Chen:2018hqy}.

For the $B_s-\bar{B}_s$ mixing, in our scenario, the dominant NP contribution to the mass difference $\Delta M_s$ comes from the one-loop box diagrams involving the charged Higgs boson exchange, and only the Wilson coefficient $C_1$ is significantly affected (see the text). The total NP contributions to $C_1$ at the initial scale $\mu_H=\mathcal{O}(M_{H^\pm})$ can be written as
\begin{equation}
C_1^{\rm NP}(\mu_H)=C_1^{HH}(\mu_H)+C_1^{HW(G)}(\mu_H).
\end{equation}
The part of $C_1^{\rm NP}(\mu_H)$ from the pure charged Higgs boson box diagrams can be written as
\begin{align}
C_1^{HH}(\mu_H)&=-\dfrac{4 v^4}{M_W^2}\bigg\{\vert\epsilon_{tc}\vert^4 D_{00}(m_c^2,m_c^2,M_{H^\pm}^2,M_{H^\pm}^2)
\bigg.
\nonumber \\
&+  \bigg. \dfrac{V_{cs}^2}{V_{ts}^2}\epsilon_{tt}^2\epsilon_{ct}^{*2}D_{00}(m_t^2,m_t^2,M_{H^\pm}^2,M_{H^\pm}^2)
\bigg.
\nonumber \\
&+  \bigg.\dfrac{V_{cs}}{V_{ts}}\epsilon_{tt}\epsilon_{ct}^*\vert \epsilon_{tc}\vert^2\Big[D_{00}(m_t^2,m_c^2,M_{H^\pm}^2,M_{H^\pm}^2)
\Big.  \bigg.
\nonumber \\
&+ \bigg. \Big. D_{00}(m_c^2,m_t^2,M_{H^\pm}^2,M_{H^\pm}^2)\Big]\bigg\},
\end{align}
while the part from the $H^\pm$-$W^\pm$ and $H^\pm$-$G^\pm$ box diagrams is given by
\begin{align}
C_1^{HW(G)}(\mu_H)&=\dfrac{8 v^2 m_t^2}{M_W^2}\left(\dfrac{V_{cs}}{V_{ts}}\epsilon_{ct}^* \epsilon_{tt}+\vert \epsilon_{tt}\vert^2\right)
\nonumber \\
&\times \Big[M_W^2 D_0(M_W^2,M_{H^\pm}^2,m_t^2,m_t^2)
\Big.
\nonumber \\
&\qquad - \Big. D_{00}(M_W^2,M_{H^\pm}^2,m_t^2,m_t^2)\Big],
\end{align}
where $D_{00}(m_1^2,m_2^2,m_3^2,m_4^2)$ and $D_0(m_1^2,m_2^2,m_3^2,m_4^2)$ correspond to the Passarino-Veltman functions~\cite{Passarino:1978jh} in the \texttt{FeynCalc} convention~\cite{Mertig:1990an,Shtabovenko:2016sxi}. Note that the result of $C_1^{\rm NP}(\mu_H)$ in our scenario can be obtained by adjusting the relevant Yukawa couplings in Ref.~\cite{Crivellin:2013wna}, which is also consistent with the ones used in Refs.~\cite{Altunkaynak:2015twa,Iguro:2018qzf,Chen:2018hqy}.

\bibliographystyle{apsrev4-1}
\bibliography{reference}

\end{document}